\documentclass[sigconf,10pt]{acmart}
\renewcommand\footnotetextcopyrightpermission[1]{} %

\usepackage{balance}
  
\usepackage{booktabs} %
\usepackage{cleveref}
\usepackage{url}

\graphicspath{{./figures/}}
\usepackage{graphicx}
\usepackage{subcaption}

\usepackage{xspace}
\newcommand*{\eg}{e.g.,\@\xspace}

\usepackage{tikz}
\usetikzlibrary{shapes,arrows,calc}
\definecolor{color1Bright}{HTML}{00bfa5}
\tikzstyle{alexNumber} = [{pos=0.5, text centered, font={\sffamily}, draw, circle, fill=color1Bright, text=white, inner sep=1}]

\def\BibTeX{{\rm B\kern-.05em{\sc i\kern-.025em b}\kern-.08emT\kern-.1667em\lower.7ex\hbox{E}\kern-.125emX}}

\begin{document}

\title{Estimating Cardinalities with Deep Sketches}

\author{Andreas Kipf}
\affiliation{\institution{Technical University of Munich}}
\email{kipf@in.tum.de}

\author{Dimitri Vorona}
\affiliation{\institution{Technical University of Munich}}
\email{vorona@in.tum.de}

\author{Jonas M\"uller}
\affiliation{\institution{Technical University of Munich}}
\email{jonas.mueller@in.tum.de}

\author{Thomas Kipf}
\affiliation{\institution{University of Amsterdam}}
\email{t.n.kipf@uva.nl}

\author{Bernhard Radke}
\affiliation{\institution{Technical University of Munich}}
\email{radke@in.tum.de}

\author{Viktor Leis}
\orcid{0000-0001-5676-8017}
\affiliation{\institution{Technical University of Munich}}
\email{leis@in.tum.de}

\author{Peter Boncz}
\affiliation{\institution{Centrum Wiskunde \& Informatica}}
\email{boncz@cwi.nl}

\author{Thomas Neumann}
\affiliation{\institution{Technical University of Munich}}
\email{neumann@in.tum.de}

\author{Alfons Kemper}
\affiliation{\institution{Technical University of Munich}}
\email{kemper@in.tum.de}

\thanks{{\bf Acknowledgements.} T.K. acknowledges funding by SAP SE}

\begin{abstract}
We introduce Deep Sketches, which are compact models of databases that allow us to estimate the result sizes of SQL queries.
Deep Sketches are powered by a new deep learning approach to cardinality estimation that can capture correlations between columns, even across tables.
Our demonstration allows users to define such sketches on the TPC-H and IMDb datasets, monitor the training process, and run ad-hoc queries against trained sketches.
We also estimate query cardinalities with HyPer and PostgreSQL to visualize the gains over traditional cardinality estimators.
\end{abstract}

\maketitle

\section{Introduction}

We introduce Deep Sketches, compact model-based representations of databases that allow us to estimate the result sizes of SQL queries.
Deep Sketches are powered by a new deep learning approach to cardinality estimation~\cite{DBLP:conf/cidr/KipfKRLBK19} (code:~\cite{learnedpy}).
This approach builds on sampling-based estimation and addresses its weaknesses when no sampled tuples qualify a predicate.
A Deep Sketch is essentially a wrapper for a (serialized) neural network and a set of materialized samples.

Estimates of intermediate query result sizes are the core ingredient to cost-based query optimizers~\cite{qoleis,leis2018query}.
While the focus of this work is \emph{not} to show the effect of better cardinality estimates on the quality of resulting query plans---which is orthogonal to having better estimates in the first place, we demonstrate that the estimates produced by Deep Sketches are superior to estimates of traditional optimizers and often close to the ground truth.
The estimates produced by Deep Sketches can directly be leveraged by existing, sophisticated join enumeration algorithms and cost models.
This is a more gradual approach than the one taken by machine learning (ML)-based end-to-end query optimizers~\cite{DBLP:conf/cidr/MarcusP19}.

Our demonstration allows users to define Deep Sketches on the TPC-H and Internet Movie Database (IMDb) datasets.
The latter is a real-world dataset that contains many correlations and therefore proves to be very challenging for cardinality estimators.
To create a Deep Sketch, users select a subset of tables and define a few parameters such as the number of training queries.
Users can then monitor the training progress, including the execution of training queries and the training of the deep learning model.
Once the model has been trained, users can issue ad-hoc queries against the resulting Deep Sketch.
Our user interface makes it easy to create such queries graphically.
Users can optionally specify a placeholder for a certain column to define a query template.
For example, a movie producer might be interested in the popularity of a certain keyword over time:
\begin{verbatim}
SELECT COUNT(*)
FROM title t, movie_keyword mk, keyword k
WHERE mk.movie_id=t.id AND mk.keyword_id=k.id
AND k.keyword='artificial-intelligence'
AND t.production_year=?
\end{verbatim}
A placeholder has a similar effect as a group-by operation, except that it does not operate on all distinct values of the group-by column but instead only on the values present in the column sample that comes with the sketch.
In other words, we instantiate the query template with values (literals) from the column sample.
Besides this being an interesting feature for data analysts, it serves the purpose of visualizing the robustness of our deep learning approach to cardinality estimation\footnote{Note that the deep learning model is not necessarily trained with literals present in the column sample. In fact, it can happen (and is even likely for columns with many distinct values) that a literal from the column sample has never been seen by the model.}.
The result of a query template can be displayed as a bar or as a line plot with one data point per template instance.
Using overlays, we show the difference to the cardinality estimates produced by HyPer\footnote{We are referring to the research version of HyPer developed at the Technical University of Munich.}~\cite{hyper} and PostgreSQL as well as to the true cardinalities---obtained by executing the queries with HyPer.

Deep Sketches feature a small footprint size (a few MiBs) and are fast to query (within milliseconds).
Due to these facts, another application of Deep Sketches lies in the area of previewing query result sizes.
Often, rough estimates are sufficient to inform users whether executing a certain query would be worthwhile, and sometimes even all they need---\eg to get an idea about the data distribution of a certain column given some selections and/or joins.
For example, Deep Sketches could be deployed in a web browser or within a cell phone to preview query results.

\section{Deep Sketches}
\label{sec:deepsketches}

\begin{figure*}
    \centering
    \begin{subfigure}[b]{0.54\textwidth}
        \includegraphics[width=\textwidth]{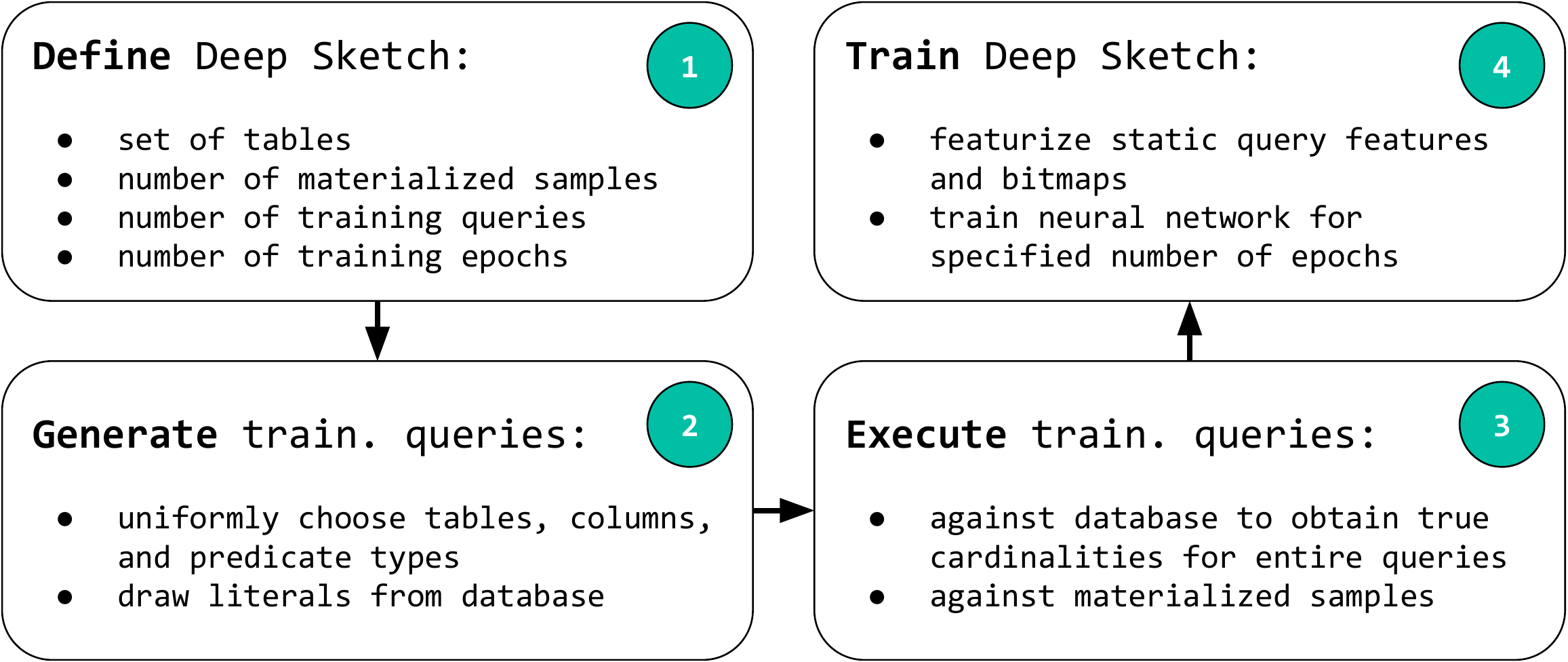}
        \caption{Creation of a sketch.}
        \label{fig:training}
    \end{subfigure}
    \begin{subfigure}[b]{0.44\textwidth}
        \includegraphics[width=\textwidth]{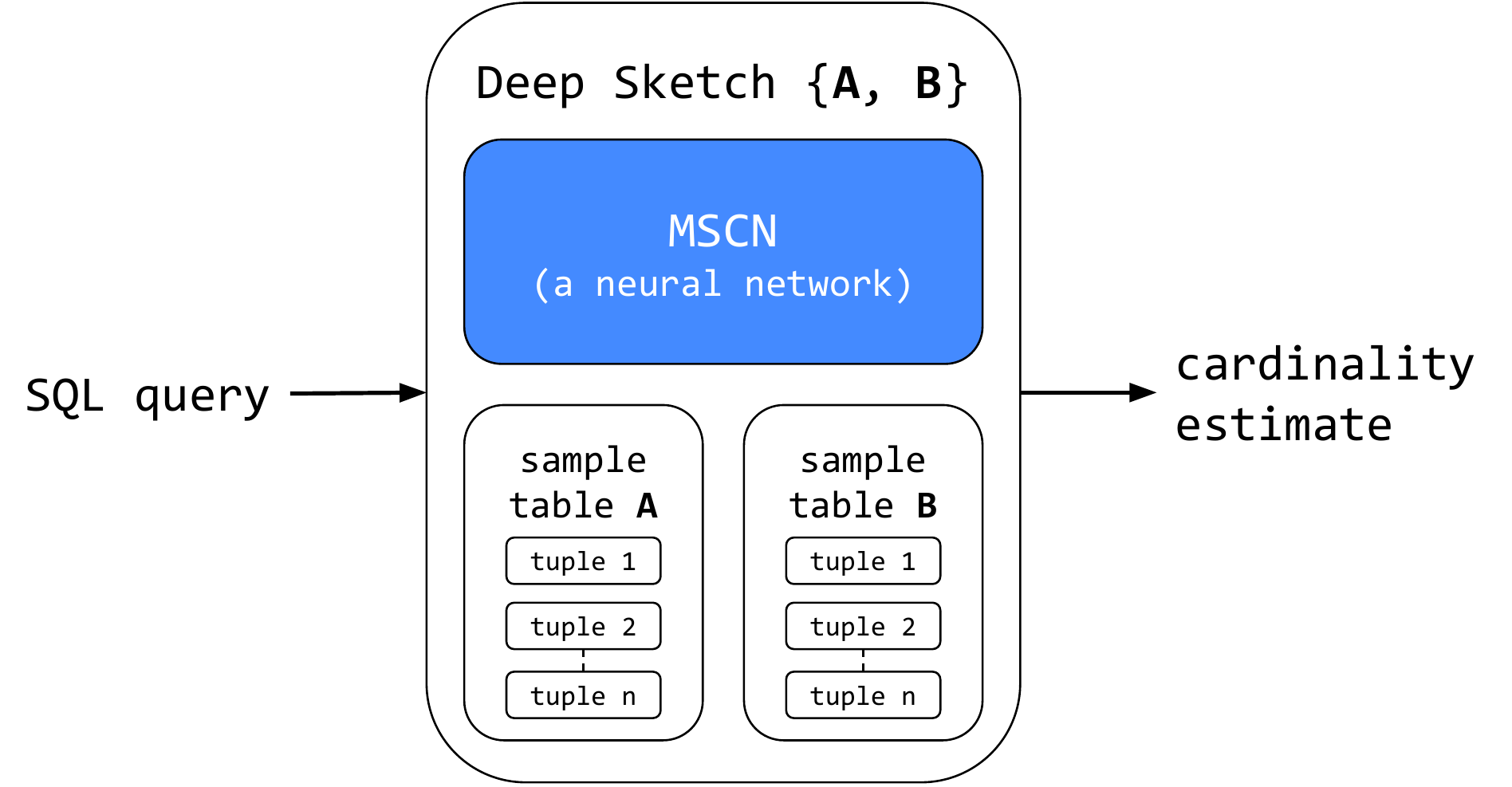}
        \caption{Result size estimation with a sketch.}
        \label{fig:inference}
    \end{subfigure}
    \caption{Creation and usage of a Deep Sketch. Depending on the number of training queries, training can be expensive. However, once a sketch is trained, it allows for an efficient result size estimation of SQL queries.}
    \label{fig:training-inference}
\end{figure*}

Deep Sketches are powered by a new (supervised) deep learning approach to cardinality estimation~\cite{DBLP:conf/cidr/KipfKRLBK19} that has recently been adopted in~\cite{DBLP:journals/corr/abs-1903-09999}.
The idea of this approach is to execute generated queries against a database to obtain true cardinalities (\emph{labels} in ML), featurize these queries, and feed them into a neural network.
Of course, instead of generating queries and uniformly sampling the search space, one could also use past user queries.

Besides static query features---such as selections or joins, we featurize information about qualifying base table samples.
In other words, in addition to executing a training query against the full database, we execute each base table selection against a set of materialized samples (\eg 1000 tuples per base table).
Thus, we derive \emph{bitmaps} indicating qualifying samples for each base table.
These bitmaps are then used as an additional input to the deep learning model.
Besides this integration of (runtime) sampling, another differentiating factor from other learning-based approaches to cardinality estimation~\cite{DBLP:conf/sigmod/OrtizBGK18, DBLP:journals/pvldb/WuJAPLQR18} is the use of a model that employs \emph{set semantics}, inspired by recent work on \textit{Deep Sets}~\cite{zaheer2017deep}, a neural network module that operates on sets.
This decision rests on the fact that the cardinality of a query is independent of its query plan---\eg both $(A \Join B) \Join C$ and $A \Join (B \Join C)$ can be represented as $\{A, B, C\}$.
While the Deep Sets model only addresses single sets, our model---called multi-set convolutional network (MSCN)---represents three sets (tables, joins, and predicates) and can capture correlations between sets.

On a high level, the MSCN model can be described as follows:
For each set, it has a separate module, comprised of one fully-connected multi-layer perceptron (MLP) per set element with shared parameters.
We average module outputs, concatenate them, and feed them into a final output MLP, which captures correlations between sets and outputs a cardinality estimate.
The featurization of a query is very straightforward.
Based on the training data, we enumerate tables, columns, joins, and predicate types ($=$, $<$, and $>$) and represent them as unique one-hot vectors.
We represent each literal in a query as a value $val$ ($val \in[0, 1]$), normalized using the minimum and maximum values of the respective column.
Similarly, we logarithmize and then normalize cardinalities (labels) using the maximum cardinality present in the training data.
For a detailed description of the model and the featurization, we refer the reader to~\cite{DBLP:conf/cidr/KipfKRLBK19}.
Finally, we train our model with the objective of minimizing the mean \emph{q-error}~\cite{DBLP:journals/pvldb/MoerkotteNS09} $q$ ($q \geq 1$).
The q-error is the factor between the true and the estimated cardinality.

One advantage of our approach over pure sampling-based cardinality estimators is that it addresses 0-tuple situations, which is when no sampled tuples qualify.
In such situations, sampling-based approaches usually fall back to an ``educated'' guess---causing large estimation errors.
Our approach, in contrast, handles such situations reasonably well \cite{DBLP:conf/cidr/KipfKRLBK19} as it can use the signal of individual query features (\eg predicate types) to provide a more precise estimate.
In addition, it can---to some degree---capture correlations across joins and can thus estimate joins without assuming independence.

\begin{table}
  \small
\begin{tabular}{@{}lllllll@{}}
\toprule
                 & median        & 90th          & 95th          & 99th          & max           & mean          \\ \midrule
Deep Sketch      & \textbf{3.82}          & \textbf{78.4} & \textbf{362} & \textbf{927} & \textbf{1110} & \textbf{57.9}   \\
HyPer            & 14.6          & 454           & 1208          & 2764          & 4228          & 224           \\
PostgreSQL       & 7.93          & 164           & 1104          & 2912          & 3477          & 174           \\ \bottomrule
\end{tabular}
\caption{Estimation errors on the JOB-light workload.}
\label{tab:job-qerror}
\vspace{-1.0em}
\end{table}

Table~\ref{tab:job-qerror} shows the estimation errors (q-errors) of our approach (Deep Sketch) compared to the cardinality estimators of HyPer and PostgreSQL version 10.3.
The results are on JOB-light~\cite{learnedpy}, which is a workload derived from the Join Order Benchmark (JOB)~\cite{qoleis} containing 70 of the original 113 queries.
In contrast to JOB, JOB-light does not contain any predicates on strings nor disjunctions and only contains queries with one to four joins.
Most queries in JOB-light have equality predicates on dimension table attributes. The only range predicate is on \texttt{production\_year}.
Considering that MSCN was trained with a uniform distribution between $=$, $<$, and $>$ predicates, it performs reasonably well.
This experiment shows that MSCN can generalize to workloads with distributions different from the training data.
By allowing users of our demonstration to issue ad-hoc queries, we want to enable them to experience this generalizability.

We believe that Deep Sketches are an important step towards a learned database system~\cite{DBLP:conf/cidr/KraskaABCKLMMN19}, and can be used in conjunction with other recently proposed ML-powered components for join enumeration~\cite{DBLP:conf/sigmod/MarcusP18, DBLP:journals/corr/abs-1808-03196}, adaptive query processing~\cite{DBLP:journals/pvldb/TrummerMMJA18}, indexing~\cite{DBLP:conf/sigmod/KraskaBCDP18, DBLP:journals/corr/abs-1903-11203}, view materialization~\cite{DBLP:journals/corr/abs-1903-01363}, workload management~\cite{DBLP:conf/cidr/JainYCH19}, and query performance prediction~\cite{DBLP:journals/corr/abs-1902-00132}.

\section{Demonstration}
\label{sec:demonstration}

As stated earlier, a Deep Sketch is essentially a wrapper for a trained MSCN model and a set of materialized samples.
In our demonstration, users can experience the end-to-end process of defining, training, and using trained sketches to estimate the result sizes of ad-hoc SQL queries.
We support the TPC-H and IMDb datasets.

Figure~\ref{fig:training} shows the four steps involved to create a new sketch.
First (\tikz[baseline=-0.7ex]{\node[alexNumber] {1}}), users need to select a subset of tables from either schema and define a few parameters, including the number of materialized base table samples, the number of training queries, and the number of training epochs.
Next (\tikz[baseline=-0.7ex]{\node[alexNumber] {2}}), we generate uniformly distributed training queries on the specified tables in our backend, and (\tikz[baseline=-0.7ex]{\node[alexNumber] {3}}) execute them with HyPer to obtain true cardinalities and to extract bitmaps indicating qualifying samples.
To accelerate this process during our demonstration, we plan to execute the training queries (in parallel) on multiple HyPer instances.
Finally (\tikz[baseline=-0.7ex]{\node[alexNumber] {4}}), we featurize the training queries and train the MSCN model for the specified number of epochs.

To give a point of reference on the training costs, training the model with 90,000 queries over 100 epochs takes almost 39 minutes on an Amazon Web Services (AWS) ml.p2.xlarge instance using the PyTorch framework~\cite{pytorch} with CUDA.
Since this number is too high for an interactive user experience, we address this problem in three ways.

First, we allow users to control the number of training queries and epochs.
For a small number of tables, 10,000 queries will already be sufficient to achieve good results.
Note that the training time decreases linearly with fewer epochs.
From our experience, 25 epochs are usually enough to achieve a reasonable mean q-error on a separate validation set.
Second, we offer pre-built (high quality) models that can be queried right away.
Third, we allow users to train new models \emph{while} querying existing ones.

Figure~\ref{fig:inference} illustrates a Deep Sketch on two tables A and B.
The interface of a sketch is very simple, it consumes a SQL query and returns a cardinality estimate.

\begin{figure*}
\centering
\includegraphics[width=0.55\linewidth]{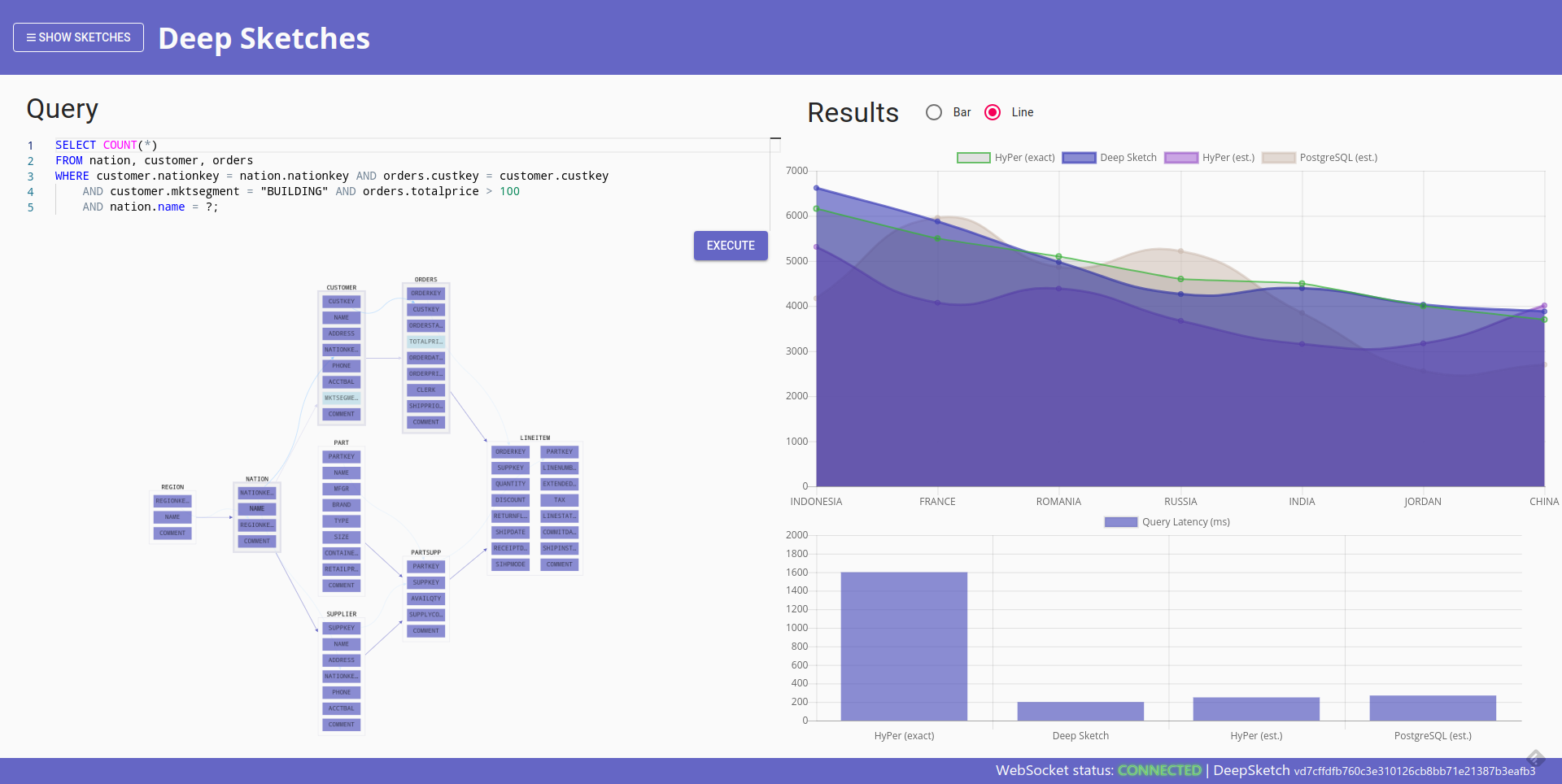}
\caption{Web interface for Deep Sketches.}
\label{fig:deepsketches}
\end{figure*}

Figure~\ref{fig:deepsketches} shows our web interface for Deep Sketches.
On the left, we allow users to specify SQL queries, and on the right, we display query results (estimated and true cardinalities).
On the top (\texttt{SHOW SKETCHES}), users can select existing and create new sketches as described above.
To query a sketch, users do not need enter SQL directly (the SQL string shown in the web interface is only displayed for information purposes).
Instead, we provide them with a simple graphical query interface.
By clicking on a table, it is added to the query.
When a user selects multiple tables, we automatically add the corresponding join predicates to the query (based on the single PK/FK relationships that exist between tables).

Users can also define selections on base tables by clicking the respective columns in the schema.
We support both equality and range predicates.
In addition, we allow users to specify a placeholder for a certain column.
Since our Deep Sketch implementation can only estimate single queries, we automatically instantiate such query templates and---in the background---execute each instance separately against the sketch.
To create such an instance, we draw a value from the column sample that is part of the sketch.
Optionally, users can select a \emph{function} to be applied to these values.
For example, for columns with many distinct values---such as \texttt{Date} columns, users may want to ``group'' the results by year (\eg \texttt{EXTRACT(YEAR FROM date)}).
To serve such queries, we generate multiple range queries (one for each year found in the sample) to be issued against the sketch.
We also support grouping the output into equally sized buckets based on the minimum and maximum values from the sample.

When a user hits the \texttt{EXECUTE} button, we issue the query against HyPer to compute its true cardinality as well as against the Deep Sketch and the cardinality estimators of HyPer and PostgreSQL to obtain estimates.
The query results are displayed with different overlays as they arrive.
On the X-axis we denote values from the placeholder column and on the Y-axis we plot the estimated and true cardinalities.
We support both bar and line charts and allow users to hide the results of individual systems.

We also use TensorBoard~\cite{tensorboard} to visualize the neural network architecture of our model and the training phase.

\section{Conclusions}
\label{sec:conclusions}

We have introduced Deep Sketches, which are compact representations of databases that allow us to estimate the result sizes of SQL queries.
Our demonstration allows users to experience the end-to-end training and querying process of these sketches.
The goal of this work is to show that a learned cardinality model can compete with and even outperform traditional cardinality estimators, especially for highly correlated data.
Our audience can specify ad-hoc queries and thereby observe that ML might indeed be the right hammer for the decades-old cardinality estimation job.
Clearly, more research is needed to automate the training and utilization of Deep Sketches in query optimizers.
One question---that we currently outsource to our users---is for which schema parts we should build such sketches.
Besides further improving cardinality estimation for query optimization, another avenue we could take in future work is to deploy Deep Sketches---which offer a small footprint size---on the client to preview query results.
\balance

\bibliographystyle{abbrv}
\bibliography{main}

\end{document}